\documentclass[twocolumn,showpacs,preprintnumbers,amsmath,amssymb,epsfig,widetext]{revtex4}

\usepackage{graphicx}
\usepackage{dcolumn}
\usepackage{bm}
\usepackage{epsfig}

\def\fun#1#2{\lower3.6pt\vbox{\baselineskip0pt\lineskip.9pt
  \ialign{$\mathsurround=0pt#1\hfil##\hfil$\crcr#2\crcr\sim\crcr}}}

\input epsf

\def\be{\begin{equation}}
\def\ee{\end{equation}}
\def\ba{\begin{eqnarray}}
\def\ea{\end{eqnarray}}

\def\nn{\nonumber}

\begin{document}

\preprint{}

\title{A Cosmological Test of Standard Gravity by Weak Lensing}

\author{Yong-Seon Song}
\email{ysong@cfcp.uchicago.edu}
\affiliation{Kavli Institute for
Cosmological Physics, Department of Astronomy \& Astrophysics, and
Enrico Fermi Institute, University of Chicago, Chicago IL 60637 }


\begin{abstract}
The large scale structure formation
in the standard Einstein gravity at later time
is uniquely determined by the expansion history $H(z)$
measured by the geometrical factor.
A possible departure from the expected standard structure formation
is detectable by growth structure test.
We design new deviation parameters to test standard gravity at large scales
and estimate the expected errors by weak lensing experiment.
The effective Newtonian constant $G_N$ at the present time can be constrained
around 10$\%$ accuracy 
by measuring the difference curvature perturbations $(\Phi-\Psi)/2$
within 2$\%$ accuracy.
\end{abstract}



\maketitle


\noindent {\it Introduction}
The gravity known to us by direct measurement ranges just 
from mm to the solar system scales.
Considering the cosmological gravitational effects beyond those scales,
we are still free to imagine a possible broken standard gravity.
The observed cosmic acceleration~\cite{perlmutter98}
can be the evidence of a departure from standard gravity at large scales.
With the presence of standard gravity, 
cosmic acceleration can be explained by introducing an exotic energy
component like dark energy (DE).
With the absence of standard gravity,
cosmic acceleration can be generated by modified gravity (MG).
The extra dimension can be introduced to generate cosmic acceleration
on a 4D brane world (DGP)~\cite{dvali00,deffayet00}.
Or we can modify the Ricci scalar $R$ with correction terms
which are proportional to the inverse of $R$~\cite{carroll03}.
Cosmic acceleration is probed by the gravitational effects
like the geometrical factor (GEF) and the growth factor (GRF).
By using the simultaneous fitting GEF and GRF,
we are able to distinguish 
DE and MG~\cite{lue04,song04,schimd04,knox05,ishak05,linder05,sawicki05,koyama05}.

In this letter, we pay attention to test standard gravity at large scales.
In standard gravity paradigm,
observable GRF by weak lensing
is uniquely determined 
by measured GEF precisely.
The expected fixed evolution of GRF leads us to think the way
to detect the deviation from standard gravity.
Inspired by the DGP type MG models,
we design deviation parameters to probe that departure.
We add the deviation parameters to the usual DE cosmological
parameters.
If we find the fixed GRF expected marginally by GEF measurement,
then we can claim the evidence of Einstein gravity at largest scales.
But if we find any significant departure from the expected DE GRF,
then we can argue that standard gravity is broken down at large scales
(not necessarily excluding DE).

We estimate the expected deviation parameter errors
by using cosmic shear experiments.
The shapes of galaxies are distorted by the intervening lensing
potential formed by GRF and projected 
by GEF.
Thus it is a best combined test of GEF and GRF.
The future weak lensing experiment like G2$\pi$
puts a tight constraint on GRF within 2$\%$ accuracy
which leads to 
the precise test of standard gravity at large scales.
We find that the effective Newtonian constant $G_N$
can be constrained around 10$\%$ level of precision.

\noindent {\it GEF test}
We fit DGP type MG models to find out if GEF test alone can 
test standard gravity broken down.
The combined test of SN and CMB is considered here.
We use the SN gold set~\cite{riess04} and the angular diameter distance 
$D(z_{\rm lss})$ to the last scattering surface measured 
by WMAP~\cite{spergel03}.
We fix the distance to
recombination at $z_{\rm lss} = 1088^{+1}_{-2}$ calculated by
the acoustic peak scale $\ell_A = 299\pm 2$ and its length calibration
through the matter density $\Omega_{\rm m} h^2=0.146\pm 0.01$.

To fit SN at $z<2$, DGP needs low matter density around $\Omega_m\sim 0.2$
which is 60$\%$ lower than best fit $\Omega_m$ of DE.
But with such a low $\Omega_m$, we are not able to get the longer 
$D(z_{\rm lss})$ measured by WMAP.
It leads $\Delta\chi^2\sim 5$ worse fit for DGP models in comparison 
with DE.
Flat DGP models fail to the combined test of SN and CMB at
around the $2-\sigma$ confidence level.

Though flat DGP type MG model fails to pass the test of SN and CMB,
relaxing the flat constraint helps DGP to fit dataset~\cite{fairbairn05}.
By adding curvature, we are able to elongate $D(z_{\rm lss})$
with keeping low $\Omega_m$.
If we put a few $\%$ curvature density to the total energy density,
we have well-matched $D(z_{\rm lss})$.
For example, the curvature with $\Omega_k=0.05$ gives DGP almost
identical $\chi^2\sim 178$ of DE.
But we do need a higher $H_0$ for DGP around 80km/s/Mpc 
which is $1-\sigma$ level off from the Hubble constant measurement
reported by HST~\cite{freedman00}.
It is also possible that any other independent measurement of 
$\Omega_k$~\cite{bernstein05}
or $H_0$ can also discriminate DE and DGP in the future.

Inspired by DGP brane world model, the DGP type MG models can be
extended generally~\cite{dvali03a}.
The modified Friedman equation can be written in the following ways
\ba
H^2-\frac{H^{2/n}}{r_c^{2-2/n}}=\frac{8\pi G}{3}\rho
\ea
where the crossover distance $r_\text{c}$
is defined as the ratio of 5-dimensional to
4-dimensional Planck mass scales
\ba
r_\text{c}=\frac{M^{(4)\,2}_\text{Pl}}{2M^{(5)\,3}_\text{Pl}}.
\ea
The crossover distance $r_\text{c}$ is a free parameter of DGP. 
If $r_\text{c}$ is close to
the current horizon scale, the acceleration of cosmic expansion is
replicated without dark energy.
If $n=2$, we have known DGP theory and if $n=\infty$, then we have the
cosmological constant.
We find that DGP type MG models with $n\ge 2$ shows good fit to 
SN+CMB.
GEF test alone shows that the current GEF measurement
does not rule out any theory.
We start our estimation based upon the expected failure of test
by GEF alone.

\noindent {\it Deviation parameters}
The combined test of GEF and GRF leads to the best opportunity
to test standard gravity at large scales.
The GEF and GRF can be simultaneously determined at the weak gravitational
lensing maps distorting the shape of galaxies systematically.
The transverse shear $\gamma_t$ of the shapes of galaxies 
is gravitationally lensed by the intervening mass.
The statistically measurable quantity $\gamma_t$
is generated by the GEF projection of the transverse displacement
driven by GRF.

We group galaxies in $n_b$ redshift bins and define the tangential shear
$\gamma_i$ in given bin $i$.
The two point function of this shear, $\langle \gamma_i\gamma_j \rangle$,
includes the information of both GEF and GRF.
The power spectra in the spherical harmonic space $C_{\sc l}$
are used in our statistical estimation, defined by
$\langle \gamma_i^{{\sc l}m}\gamma_j^{{\sc l}'m'} \rangle 
=C_{\sc l}\delta_{{\sc l}{\sc l}'}\delta_{mm'}$.
We have $n_b(n_b+1)/2$ shear power spectra which are given by~\cite{hu99a,takada03,song03}
\ba
C_l^{ij} = \pi^2 {l\over 2} \int dr r W^i(r) W^j(r)\Delta_-^2(k,z(r))
\ea
where $W(\bar r_i,r)$ is the weight function~\cite{song03} and
the dimensionless power spectrum $\Delta_-$ is two point function
of the quantity related to the net tangential deflection.

The net tangential displacement by the gravitational lensing
is given by the spatial gradient of
the difference curvature perturbations 
$\Phi_-=(\Phi-\Psi)/2$, which is written as in real space 
(without $(r)$, $\Phi$ and $\Psi$ are in Fourier space.)
\ba
\nabla_{\perp}\Phi_-(r)=\nabla_{\perp}[\Phi(r)-\Psi(r)]/2
\ea
In the standard gravity frame,
the difference curvature perturbation $\Phi_-$ is equal to $\Phi$
with the condition $\Phi_+=(\Phi+\Psi)/2=0$
of ignorable anisotropy stress.
The power spectrum $\Delta_-$ is given by
\ba
\Delta_-={k^3 \over 2\pi^2}\langle \Phi_- \Phi_-\rangle.
\ea
is not different from $\Delta_{\Phi}$ in standard gravity.
But in general MG models,
$\Delta_-$ is not always equal to $\Delta_{\Phi}$.
What we measure by the weak lensing is the difference curvature
perturbations $\Phi_-$.

We need three steps to calculate the measurable quantity $\Phi_-$:
1) to derive the matter perturbations,
2) to relate the matter perturbations to the curvature perturbations
through non-dynamical constraint equation given by 00-component
of Einstein equations and
3) to find the anisotropy sources which break the condition of 
$\Phi_+=0$.
We derive the matter perturbations by solving the following 
differential equations,
\ba\label{gm}
\ddot{\delta}_{\rm m}+\frac{H}{1+z}\dot{\delta}_{\rm m}=-k^2\Psi,
\ea
where the evolution of $\delta_{\rm m}$ is sourced by
the Newtonian constant $\Psi$. Next the curvature perturbations
$\Phi$ is related to $\delta_{\rm m}$ by Poisson equation given by
\ba\label{Po}
k^2\Phi=4\pi Ga^2\rho_m\delta_m.
\ea
Then we get $\Phi_-$ from the given condition $\Phi_+$.

In DE model (standard gravity case),
there is no modification to the basic form of Eq.~\ref{gm} and Eq.~\ref{Po}.
The expansion history $H(z)$ measured by GEF completely fixes $\Phi_-$.
In MG model, the modified $\Psi$ affects on the evolution of $\delta_{\rm m}$
differently from standard gravity. Also the relation 
between $\Phi$ and $\delta_{\rm m}$ is altered by the modified Poisson 
equation.
We would get the distinct $\Phi_-$ of MG through
possible modifications on Eq.~\ref{gm}, Eq.~\ref{Po}
and the modified condition $\Phi_+$.

The general behavior of the deviation of $\Phi_-$ evolution
from standard gravity can be deduced from the DGP type
modification.
In the DGP type model, the metric perturbations are modified
in scalar-tensor type
\ba
k^2\Phi&=&4\pi a^2G\left[1+\beta_{\Phi}^{\text DGP}(H)\right]\rho_m\delta_m\nn\\
k^2\Psi&=-&4\pi a^2G\left[1+\beta_{\Psi}^{\text DGP}(H)\right]\rho_m\delta_m
\ea
Though we do not have the full solution of the 5 dimensional perturbation 
theory, we solve the 4D gauge theory on the brane based on the ansatz
which closes the 4D Bianchi identity.
The exact form of 
$\beta_{\Phi}^{\text DGP}(H)$ and $\beta_{\Psi}^{\text DGP}(H)$
depends on our ansatz on the 4D brane.

\begin{figure}[htbp]
  \begin{center}
  \epsfysize=5.truein
  \epsfxsize=3.3truein
    \epsffile{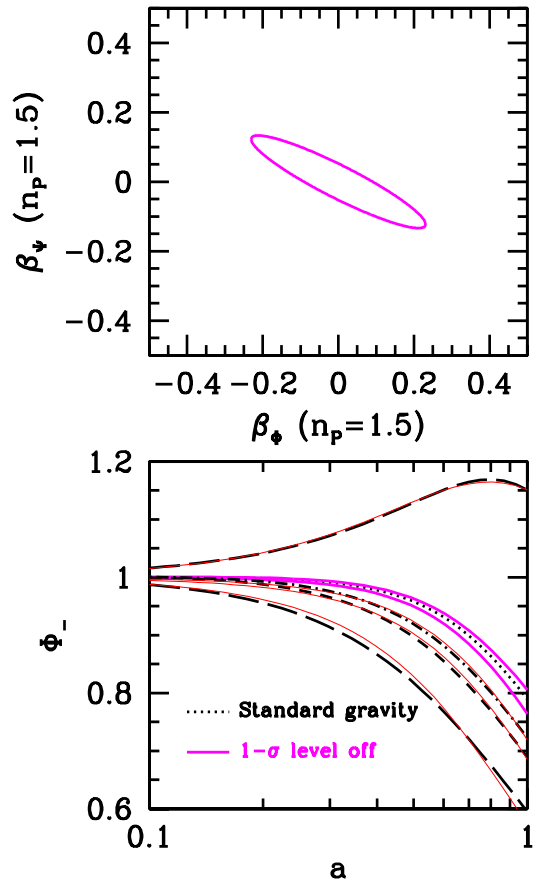}
    \caption{\footnotesize 
{\bf Top panel}: the expected deviation parameter errors
of $\beta_{\Phi}$ and $\beta_{\Psi}$ 
calculated by G2$\pi$ weak lensing survey at fixed $n_P=1.5$.\\
\noindent {\bf Bottom panel}: 
we show the observable structure formation by weak lensing
$\Phi_-$.
Dotted line-standard gravity; 
two thick solid lines surrounding the standard gravity
solution-translating $1-\sigma$ off curves of the top panel 
in $\Phi_-$ space; 
two long dash lines-$n=2$ case, upper one for 4D DGP and lower one for
static DGP;
a dash lines and a dot dash line-$n=3$ and $n=4$ cases.
We show the reconstructed $\Phi_-$ by our beta parameters
at fixed $n_P=1.5$. Four thin solid lines denote the reconstructed
$\Phi_-$ which show great fit to actual theories.
}
\label{fig:1}
\end{center}
\end{figure}

In 4D DGP, we assume no significant contribution from
the 5th dimensional projected Weyl fluid~\cite{song04,sawicki05}. 
This ansatz leads to no anisotropy stress on the brane,
i.e. the condition $\Phi_+=0$ still holds
\ba
\beta_{\Phi}^{\text 4D}(H)=\beta_{\Psi}^{\text 4D}(H).
\ea
The reduced fundamental mass scale results in 
the increment of $G_N$ variation which makes
both $\Phi$ and $\Psi$ increase.
Shown in Fig~\ref{fig:1} (upper long dash line in the bottom panel), 
$\Phi_-$ of 4D
gets peculiar evolution at later time
until $\Phi_-$ starts to decrease by the decay of $\delta_m$.
This ansatz does not achieve the self-consistency.
To neglect 5th dimensional effects causes the unstable
structure formation on the brane~\cite{koyama05}.
But we still use this case to show the generality of our parameterization
as a possible interesting case of MG which can boost $\Phi_-$ higher.

In static DGP, the time derivative to the spatial derivative is ignored,
which is a sound assumption for cosmological structure formation.
Then the non-negligible anisotropy stress tensor should be introduced~\cite{koyama05}.
The modification on the geometrical perturbations looks different from 
the 4D DGP case as
\ba
\beta_{\Phi}^{\text static}(H)=-\beta_{\Psi}^{\text static}(H).
\ea
While $\Phi_+=0$ condition is broken by the significant contribution
of the projected Weyl tensor,
the deflected light path by the gravitational lensing is unmodified.
Shown in Fig~\ref{fig:1} (lower long dash line in the bottom panel), 
$\Phi_-$ gets the extra decrement
in comparison with standard gravity case.
Though $\Phi_-$ experience no modification,
the weaker $\Psi$ sourcing matter perturbation leads to
decay of $\Phi_-$.

Though DGP is found to have trouble in providing the stable
structure formation, both opposite MG models help us to guess
the general behavior of scalar-tensor type modification on $\Phi_-$.
As far as the scale dependence of GRF can be separated from 
the time dependence of GRF,
we can design new deviation parameters by considering 
the modification effects on $\Phi_+$ and $\Phi_-$ like the DGP type.

It leads us to consider the following empirical parameterization as
\ba
\Phi'=(1+\beta_{\Phi}a^{\text n_p})\Phi \nn\\
\Psi'=(1+\beta_{\Psi}a^{\text n_p})\Psi
\ea
where $\beta_{\Phi}$ and $\beta_{\Psi}$ denote the constant strength of 
the deviation and $\text n_p$ determines
the moment of the deviation from standard gravity.
The shifted metric perturbations $\Phi'$ and $\Psi'$
well reconstruct the actual
shape of observable $\Phi_-$ of each theory.

The power $\text n_p$ needs to be positive to satisfy with the request of
negligible modification effect at the matter dominated epoch.
There is significant degeneracy to generate 
$\Phi_-$ among deviation parameters.
As far as $\text n_p$ is set to be reasonable value between 
$0.5 < \text n_p < 2.5$, we are able to find a different pair of 
$\beta_{\Phi}$ and $\beta_{\Psi}$ to provide the approximately 
identical $\Phi_-$ in the end.
We choose $\text n_p=1.5$ as a fixed constant and float 
$\beta_{\Phi}$ and $\beta_{\Psi}$ parameters.

We produce $\Phi_-$ based upon the real MG theory described above.
Though we use the 
nearly identical input GEF of each theory,
the difference curvature perturbations $\Phi_-$ of each theory
are noticeably different from each other
(the bottom panel of Fig~\ref{fig:1}).
There are clear distinct between 4D DGP and static DGP 
(upper long dash and lower long dash lines respectively).
The index $n$ can denote the type of newly introduced non-linear
energy momentum tensor; quadrature for $n=2$ (DGP), cubic for $n=3$
and quartic for $n=4$.
The bottom panel shows that as we increase $n$
we would have closer $\Phi_-$ of LCDM.
We get $\Phi_-$ of $n=3,4$ by using static ansatz only~\cite{koyama06}
(dash and dash-dotted lines respectively).

We try to reconstruct modified $\Phi_-$ with $\beta$ parameters
without assuming any peculiar MG theory to begin with.
The thin solid lines in the bottom panel of Fig~\ref{fig:1}
present $\Phi_-$ provided by $\beta$ parameters.
We find that our deviation parameters show great fit to any known MG theory.

\noindent {\it Forecast}
We float all known cosmological parameters without any prior information.
Those are simultaneously determined by 
shear correlation functions and CMB power spectra ({\it Planck}).
We take the same set of QCDM cosmological parameters in~\cite{song03} 
and add our $\beta$ parameters with fixed $n_P=1.5$.
We consider a reference shear survey, 
ground-based (G2$\pi$~\cite{tyson02}).
The ground-based surveys can cover more sky.  
G2$\pi$ represents the ground-based survey with a half-sky coverage.

We calculate the uncertainties of parameters by using the first order
Taylor expansion of the parameter dependence of two-point 
correlation functions.
If the assumption of this linear response is available,
the inverse of the Fisher matrix leads to the expected parameter errors
with the given expected experimental errors on the power spectra.
The contribution to the Fisher matrix from the shear-shear correlations
is given by
\ba
F^{\gamma\gamma}_{pp'}
 = \sum_{l,i1,i2,i3,i4} {2l+1 \over 2} C_{l,p}^{i1,i2} {\cal W}_l^{i2,i3}
C_{l,p'}^{i3,i4} {\cal W}_l^{i4,i1},
\ea
where the subscript $,p$ denotes differentiation with respect to parameter
$a_p$.  We use ${\cal W}$ to denote the inverse of the total covariance
matrix:
\be
{\cal W} \equiv ({\bf S} + {\bf N})^{-1}.
\ee
Note that ${\cal W}$ is an easily invertible block diagonal matrix: 
${\cal W}_{lmi,l'm'i'}={\cal W}_l^{ii'}
\delta_{mm'}\delta_{ii'}$.  The total Fisher matrix
is given by summing $F^{\gamma\gamma}$ with the Fisher matrix for the
unlensed CMB data, $F^{\rm CMB}$.

We take the standard gravity model (DE) as our fiducial one.
If DE is a righteous modeling for cosmic acceleration,
then we would be interested in the upper bound of standard gravity.
On the other hand, if MG is a correct one to explain cosmic acceleration,
then we need to know if proper MG model can be discriminated from the standard
case and from the other MG models.

Standard gravity points (0,0) on the $\beta$ space.
We estimate the expected errors with G2$\pi$ 
in the top panel of Fig~\ref{fig:1}.
The errors on $\beta$ parameters indicate the fractional difference
of modified gravitational constant from standard $G_N$ at the present time.
We get the upper bound for effective $G_N$ around 8.8$\%$ level of accuracy
with G2$\pi$ at the present time
(effective fractional error $\sigma_{G_N}$ for $\Phi_-$ is defined by 
$(\vec{\sigma_{\beta}}\cdot C^{-1}\cdot \vec{\sigma_{\beta}})^{1/2}/2$.).
The percentage error for $G_N$ ranges from 3$\%$ to 15$\%$
with $0.5 < \text n_p < 2.5$.
The estimation of $G_N$ error gives us the rough idea of constraints 
on the standard gravity.
But considering the phenomenological aspects of weak lensing,
it would be better to translate the beta parameter constraints on the space
where we can see the evolution of the structure formation.

In the bottom panel of Fig~\ref{fig:1},
we translate the constraints of beta parameters 
on $\Phi_-$ plane.
The result is really encouraging.
Approximately 10$\%$ uncertainty in $G_N$
results in about 2$\%$ level of constraints on $\Phi_-$ evolution
at later time.
If the standard gravity is true model of our universe
even at such a large scale,
then we can predict the structure formation within 2$\%$ level of accuracy
by determining GEF precisely.
It could get worse by all the systematic error expected with G2$\pi$,
but it is recommended to combine all other GEF measurements other than 
weak lensing (SN and baryon oscillation),
which will lead to much stronger constraints on $\Phi_-$ evolution
after all.

The meaning of constraints on $\Phi_-$ we get
can be seen by comparing the estimated error with the level of departure
from the standard gravity predicted by DGP type MG model
(the bottom panel of Fig~\ref{fig:1}).
Obviously both $n=2$ cases are distinguishable from the standard gravity.
Also even higher $n$ cases are marginally discriminable.
Considering all known MG models, we can conclude that
the constraints on the standard gravity
estimated by using G2$\pi$ are meaningful ones
to test the standard gravity at large scales in the future.

\noindent {\it Conclusion}
We design the deviation parameters and calculate the expected
errors of those parameters by using the combination of weak lensing and CMB.
From the expected errors on $\beta$ parameters,
we calculate the upper bound of effective $G_N(0)$ around 10$\%$ accuracy 
at the present time by measuring
$\Phi_-$ deviation from standard gravity
within 2$\%$ accuracy.
Shown in Fig~\ref{fig:1}, it is well below the level of a departure of
$\Phi_-$ reported by any known MG model so far.
DE would be strongly supported by this test
with nearly zero $\beta$ parameters measured.
On the other hand, any measurement of non-zero $\beta$ parameters
could lead us to shift to MG modeling from DE to explain 
cosmic acceleration, or
to look for any other minor broken standard gravity with
keeping DE modeling as it is.

\noindent {\it Acknowledgments}: 
We thank Christopher Gordon, Wayne Hu, Ignacy Sawicki and Xiaoming Wang
for helpful discussion.
YSS is supported by the
U.S.~Dept. of Energy contract DE-FG02-90ER-40560.
This work was carried out at the KICP under NSF PHY-0114422.


\end{document}